\documentclass[pra,onecolumn,showpacs,aps]{revtex4}
\usepackage{graphics}
\usepackage{color}
\usepackage{graphicx}
\usepackage{bm}
\newcommand{\TAD}{New Zealand Foundation for Research, Science and Technology contract TAD-1054}
\newcommand{\MF}{Marsden Fund of the Royal Society of New Zealand contract PVT-202}
\newcommand{\ARC}{Australian Research Council}

\newcommand{\GPE}{Gross-Pitaevskii equation}

\newcommand{\etal}{{\em et al.}}
\newcommand{\EQ}[1]{\begin{eqnarray}#1\end{eqnarray}}
\newcommand{\mbf}[1]{\mathbf{#1}}
\newcommand{\notes}[1]{}
\def\NOTES{
\renewcommand{\notes}[1]{\par\noindent{\color{red}\emph{##1}}\par}}
\NOTES
\begin{document}
\title{Diffusive instability of a vortex in a rotating Bose gas}
\author{A.~S. Bradley$^{1,2,}$}
\email{abradley@physics.uq.edu.au}
\author{C.~W. Gardiner$^{2,3,}$}
\email{gardiner@physics.otago.ac.nz}
\affiliation{$^1$ARC Centre of Excellence for Quantum-Atom Optics, Department of Physics, University of Queensland, Brisbane QLD 4072, Australia.}
\affiliation{$^2$School of Chemical and Physical Sciences, Victoria University of Wellington, New Zealand.}
\affiliation{$^3$Ultra-Cold Atoms Group, Department of Physics, University of Otago, Dunedin, New Zealand.}
\date{\today}
\begin{abstract}
The dissipative dynamics of a vortex in a finite temperature trapped Bose-Einstein condensate are shown to be governed by a {\em diffusive instability}. 
In the weakly interacting regime we find a cross-over from instability to metastability of the vortex, determined only by the relative rotation rate of the vortex and thermal cloud. Even in the thermodynamically stable regime there is a finite lifetime arising from diffusive instability. We find the mean exit time for the vortex from the condensate, and show that near the critical frequency for thermodynamic stability the lifetime has a universal form which is independent of the condensate size and the vortex precession frequency. In the thermodynamically stable regime the steady-state two-time position correlation predicts `thermal bunching', giving an increased probability of positional coincidence for short times.
\end{abstract}
\maketitle
\section{Introduction}
Since the achievement of atomic Bose-Einstein condensation (BEC)~\cite{Anderson1995}, the study of quantised superfluid vortices has attracted intense interest. The observation of vortices in BEC gave direct evidence of superfluidity~\cite{Madison2000,Matthews1998}, and led to studies of the dynamics of formation and decay of vortex lattices. Vortices also form a challenging testing ground of BEC theory which is renowned for its tractability, giving unique insights into quantum many-body physics.
\par
It is widely appreciated that there is an energy barrier to the penetration of a vortex into the interior of a BEC~\cite{Fetter2001}. The experimental challenge of creating condensates containing large numbers of vortices was met by generating a thermal non-condensed fraction of the gas with significant angular momentum, which acts as a rotating environment for the condensate. Stimulated collisions between condensed and thermal atoms impart angular momentum to the BEC, leading to the formation of quantised vortices. Experimental methods include stirring with an elliptical perturbing potential~\cite{Hodby2001}, with a detuned laser~\cite{Abo-Shaeer2001}, and, alternatively, surmounting the barrier {\em a priori} by evaporatively cooling a {\em rotating} thermal cloud to quantum degeneracy~\cite{Haljan2001}.
\par
Vortex motion is strongly dependent on certain types of instability~\cite{Fetter2001}, and the role of instabilities in overcoming the energy barrier has been investigated theoretically and experimentally.
 There are two well-understood forms of instability in trapped Bose-Einstein condensates, {\em dynamic} and {\em thermodynamic}~\cite{STR}. Dynamic instability arises when small perturbations of the condensate may grow with time, and is associated with an imaginary eigenfrequency of the linearised Bogoliubov-de Gennes (BdG) equations. Thermodynamic instability occurs when the system energy may reduce in the presence of dissipation, and is associated with negative eigenfrequencies. The aim of this paper is to show that the thermal motion of a vortex is strongly dependent on a new {\em diffusive instability}, and to elucidate the effect of thermal atoms on the vortex in a trapped rotating BEC.
\par 
In this work we generalise the stochastic variational method developed by Duine \etal~\cite{Duine2002,Duine2004} 
to find an effective Lagrangian for the motion of a straight line vortex in a {\em rotating} Bose gas. We verify the validity of their approach for the regime treated in Ref.~\cite{Duine2004}, and investigate the new physical processes revealed by the more general theory.
\section{The system and Vortex Action}
\begin{figure}[!htb]
\begin{center}
\includegraphics[width=.5\columnwidth]{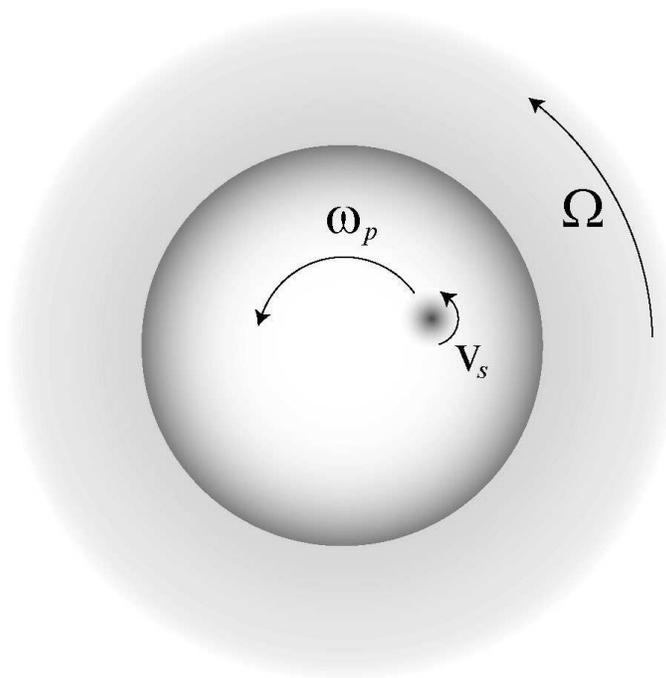}
\caption{Schematic of the system consisting of a partially condensed Bose gas in an oblate harmonic trap. The condensate contains a vortex which precesses at angular frequency $\omega_p$, and is immersed in a thermal cloud rotating at angular frequency $\Omega$. The circulation of the superfluid near the vortex core is indicated by $v_s$. All quantities are defined with respect to the laboratory reference frame.}
\label{fig:vlife_schematic}
\end{center}
\end{figure}
\subsection{Approximate high temperature description}
There have been several theories developed to descibe the dynamics of a partially condensed Bose gas. While exact theoretical methods, which in principle include all quantum coherences, are available~\cite{Steel1998,Drummond1999,Deuar2002}, practical calculations for nonlinear multimode systems using these methods are often severely limited by technical problems. The two approaches that interest us here are those of Stoof~\cite{Stoof1999,Stoof2001}, and Gardiner~\cite{SGPEI,SGPEII} who have developed closely related {\em classical} field theories of Bose-Einstein condensation in the {\em high-temperature} regime. For a mode with characteristic frequency $\omega$, the system is in the high temperature regime when the temperature satisfies
\begin{equation}
\hbar\omega \ll k_BT<k_BT_c,
\end{equation}
where $T_c$ is the temperature for Bose-Einstein condensation. The classical field approximation for the condensate mode is valid when it is macroscopically occupied, or when $1\ll N_c$. For our purposes the two theories are equivalent, so we will use the notation of Stoof to assist comparison with the paper of Duine {\em et al}~\cite{Duine2004}. 
The reader is referred to Ref.~\cite{Stoof1999} for the details of the derivation of the SGPE for the scenario where the thermal cloud is at rest in the laboratory frame; the new physical feature we include here is the possible rotation of the thermal cloud. At the level of approximation we are using the rotation may be included by minor technical modifications of the theory, which nonetheless have important physical consequences for the motion of the vortex.
\par
The system of interest consists of a partially condensed Bose gas of atoms with mass $m$ held in an axisymmetric oblate trap described by the potential 
\EQ{
V^{\rm ext}(\mbf{x})=\frac{m}{2}\left\{\omega^2(x^2+y^2)+\omega_z^2 z^2\right\},
}
where $\omega\ll \omega_z$.
In the weak interaction regime the collisions are described by the two body T-matrix, $T^{\rm 2B}=4\pi\hbar^2 a/m$, which is determined by the S-wave scattering length $a$~\cite{Dalfovo1999}.  The condensate is assumed to contain a single vortex which has a single quantum of circulation, and the thermal cloud at chemical potential $\mu$ is in a state of uniform rotation about the $z$-axis at angular frequency $\Omega$ with respect to the laboratory frame. This is shown schematically in Fig.~\ref{fig:vlife_schematic}. 
\par 
For the general case where the thermal cloud is rotating, the theory of coupling between thermal cloud and condensate is most easily formulated in the frame of reference where the cloud is stationary. This leads to some modifications of the SGPE~\cite{BradleyThesis}, arising from the transformation into the rotating frame, in the form of terms which depend on the operator for the $z$-axis component of angular momentum, $L_z$. In the frame of the thermal cloud, the essential change to the theory is to alter the single particle Hamiltonian for the Bose gas to $H_{\rm sp}\to H_{\rm sp}-\Omega L_z$~\cite{LandLSM1}. 
\par
Incorporating the physical effect of the rotating cloud into the SGPE formalism, we take our starting point as the stochastic \GPE 
\EQ{\label{SGPE}
i\hbar\frac{\partial\phi}{\partial t}&=&\left(1+\frac{\beta}{4}\Sigma^{\rm K}\right)\left\{-\frac{\hbar^2\nabla^2}{2m}+V^{\rm ext}(\mbf{x})-\Omega L_z-\mu+T^{\rm 2B}|\phi(\mbf{x},t)|^2\right\}\phi(\mbf{x},t)+\eta(\mbf{x},t),
}
where the fluctuations are completely determined by the non-zero correlator of the complex Gaussian noise amplitude
\EQ{\label{eq:noise}
\langle \eta^*(\mbf{x},t)\eta(\mbf{x}^\prime,t^\prime)\rangle=\frac{i\hbar^2\Sigma^{\rm K}(\mbf{x},t)}{2}\delta(\mbf{x}-\mbf{x}^\prime)\delta(t-t^\prime),
}
where the Keldysh self-energy arises from Bose-enhanced collisions of condensate atoms with thermal cloud atoms, and is given by~\cite{Duine2002,Duine2004,QKIII}
\EQ{
\hbar\Sigma^{\rm K}(\mbf{x},t)&=&-4\pi i(T^{\rm 2B})^2\int \frac{d\mbf{k}_1}{(2\pi)^3}\int \frac{d\mbf{k}_2}{(2\pi)^3}\int \frac{d\mbf{k}_3}{(2\pi)^3}(2\pi)^3\delta(\mbf{k}_1-\mbf{k}_2-\mbf{k}_3)\nonumber\\
&&\times\delta(\epsilon_1-\epsilon_2-\epsilon_3)[f_1(1+f_2)(1+f_3)-(1+f_1)f_2f_3],
}
where the thermal cloud is described by the rotating frame Bose-Einstein distribution within the Hartree-Fock approximation given by
\EQ{
f_i&\equiv&\left[\exp{\left(-\beta(\epsilon_i-\mu N)\right)}-1\right]^{-1},\\
\epsilon_i&=&\frac{\hbar^2\mbf{k}_i^2}{2m}+V^{\rm ext}(\mbf{x})+i\hbar\mbf{\Omega}\cdot (\mbf{x}\times\mbf{k}_i)+2T^{\rm 2B}|\langle \phi(\mbf{x},t)\rangle|^2.
}
In this work we will treat the thermal cloud as time-invariant and spatially uniform. As $|\Omega|$ increases the thermal cloud is centrifugally broadened, improving the validity of this description. We may then use the simpler approximate form~\cite{QKIII,Penckwitt2002,BradleyA}
\EQ{\label{eq:kaldysh}
\Sigma^{\rm K}=\frac{-i48ma^2(k_BT)^2}{\pi\hbar^3}.
}
Equations~\ref{SGPE}, \ref{eq:kaldysh} and \ref{eq:noise} describe a condensate driven by thermal fluctuations arising from Bose-enhanced collisions between condensate atoms and atoms in the rotating thermal cloud. The condensate is described by a modified Gross-Pitaevskii equation with appropriate damping and noise terms determined by the motion of the thermal cloud. It follows from the form of the noise and the damping terms that the SGPE automatically satisfies the fluctuation-dissipation theorem.
We remark here that the interaction of the thermal noncondensate with the condensate leads to another process, in addition to the growth terms we are concerned with here, involving collisions which do {\em not} change the condensate occupation, the so-called {\em scattering} terms~\cite{SGPEII}. However, as shown by Anglin and Zurek~\cite{Anglin1999}, in a first approximation we may neglect these terms since their net effect is to induce a small change in the effective damping and fluctuation terms in Eq.~\ref{SGPE}.
\subsection{Effective vortex action}
It has been shown in Ref.~\cite{Duine2002} that the Wigner probability distribution for the condensate wavefunction evolving according to the Langevin field equation (\ref{SGPE}) can be written as the functional integral
\EQ{
W[\phi,\phi^*]=\int_{\phi(\mbf{x},t)=\phi(\mbf{x})}^{\phi^*(\mbf{x},t)=\phi^*(\mbf{x})}d[\phi]d[\phi^*]\exp{\left\{\frac{i}{\hbar}S^{\rm eff}[\phi,\phi^*]\right\}},
}
where the effective action is
\EQ{\label{eq:Seff}
S^{\rm eff}[\phi,\phi^*]&=&\int dt^\prime\int d\mbf{x}\frac{2}{\hbar\Sigma^k}\Bigg|\Bigg(i\hbar\frac{\partial}{\partial t^\prime}+\left\{1+\frac{\beta}{4}\hbar\Sigma^k\right\}\nonumber\\
&&\times\left[\frac{\hbar^2\nabla^2}{2m}-V^{\rm ext}+\Omega L_z+\mu-T^{\rm 2B}|\phi(\mbf{x},t^\prime)|^2\right]\Bigg)\phi(\mbf{x},t)\Bigg|^2.
}
\par
In order to find the dynamics of a single vortex at finite temperature we introduce a Gaussian variational ansatz for the condensate wavefunction
\EQ{\label{eq:Gaussian}
\phi(\mbf{x},t)&=&\frac{\sqrt{N_c(t)}e^{i\theta_0(t)}}{\sqrt{\pi^{3/2}q^2q_z\left[q^2+u_x^2(t)+u_y^2(t)\right]}}\left([x-u_x(t)]+i[y-u_y(t)]\right)\nonumber\\
&&\times\exp{\left\{-\frac{x^2+y^2}{2q^2}-\frac{z^2}{2q_z^2}\right\}},
}
which describes a single line vortex in an otherwise cylindrically symmetric Bose-Einstein condensate. In the noninteracting limit the ansatz reduces to the exact wavefunction for the condensate containing a vortex, since the length parameters reduce to the harmonic oscillator lengths determined by the trap. The time dependent variational parameters $u_x(t), u_y(t)$ are the Cartesian coordinates of the vortex, and the time independent variational parameters are the condensate size parameters in the radial ($q$) and axial ($q_z$) directions. Since the condensate can exchange particles with the thermal cloud it is crucial to also allow for the time dependence of the condensate occupation, $N_c(t)$ and phase $\theta_0(t)$.
Substituting this ansatz into Eq. \ref{eq:Seff} we find the effective action
\EQ{
S^{\rm eff}[N_c,\theta_0,\mbf{u}]&=&\int_{t_0}^{t} dt^\prime\;\frac{2}{\hbar\Sigma^{\rm K}}\Bigg\{N_c(t^\prime)\left[\hbar\dot{\theta}_0(t^\prime)+\mu_c(t^\prime)-\mu+\frac{\hbar}{q^2}(u_x(t^\prime)\dot{u}_y(t^\prime)-u_y(t^\prime)\dot{u}_x(t^\prime))\right]^2\nonumber\\
&&+\frac{\hbar^2}{4N_c(t^\prime)}\left[\dot{N}_c(t^\prime)+\frac{\beta}{2}i\Sigma^{\rm K}(\mu_c(t^\prime)-\mu)N_c(t^\prime)\right]^2\nonumber\\
&&+\frac{\hbar^2N_c(t^\prime)}{q^2}[\dot{u}_x(t^\prime)+\omega_p(q,N_c)u_y(t^\prime)-\gamma u_x(t^\prime)]^2\nonumber\\
&&+\frac{\hbar^2N_c(t^\prime)}{q^2}[\dot{u}_y(t^\prime)-\omega_p(q,N_c)u_x(t^\prime)-\gamma u_y(t^\prime)]^2+{\cal O}((u_i/q)^3)\Bigg\},
}
where the damping rate, chemical potential, vortex precession frequency, and effective potential are respectively
\EQ{\label{eq:gam}
\gamma=\frac{\beta i\Sigma^{\rm K}}{4}\hbar(\omega_p(q,N_c)-\Omega),
}
\EQ{\label{eq:muc}
\mu_c(t)=\frac{\partial}{\partial N_c}N_c\left(V(q,q_z,N_c)-\hbar(\omega_p(q,N_c)-\Omega)\frac{u_x^2(t)+u_y^2(t)}{q^2}\right),
}
\EQ{
\omega_p(q,N_c)=\frac{\hbar}{2mq^2}+\frac{m\omega^2q^2}{2\hbar}-\frac{aN_c\hbar}{\sqrt{2\pi}mq^2q_z},
} 
\EQ{\label{eq:pot}
V(q,q_z,N_c)=\frac{\hbar^2}{mq^2}+m\omega^2q^2+\frac{\hbar^2}{4mq_z^2}+\frac{m\omega_z^2q_z^2}{4}+\frac{aN_c\hbar^2}{2\sqrt{2\pi}mq^2q_z}-\hbar\Omega.
}
We emphasise here that the form of the damping given by Eq.~\ref{eq:gam} depends only on the relative angular frequency of the vortex and thermal cloud. This is a simple consequence of the presence of a thermodynamic instability of the vortex when $\Omega<\omega_p$. It should also be noted that the form that occurs in Ref.~\cite{Duine2004} is obtained from Eq.~\ref{eq:gam} by setting $\Omega=\omega=a=0$, that is, only the kinetic energy contribution to $\omega_p$ was included.
\par
Thus far we have retained the time dependence of the condensate occupation and phase, and it is pleasing to see that the vortex contributes a term to $\mu_c(t)$ which depends on the relative rotation of the vortex and the cloud. To a first approximation, the change in atom number during the vortex motion may be neglected, so that in what follows we fix the phase and occupation of the condensate. The time independent variational parameters $q$ and $q_z$ are then chosen to minimise Eq.~\ref{eq:pot} for $N_c$ atoms in the condensate.
We choose to define the condensate length parameter $q$ as the edge of the condensate for the vortex motion.
\par
For a non-rotating cloud the chemical potential decreases as the square of the distance of the vortex from the center of the trap, which is consistent with the thermodynamic instability of the vortex. If the cloud rotates at the frequency of the vortex precession in the laboratory frame, $\Omega=\omega_p$,
the chemical potential becomes independent of the position of the vortex, and the vortex motion depends entirely on thermal fluctuations. When $\omega_p<\Omega$ the center of the condensate becomes a minimum of the chemical potential, increasing quadratically with the radial distance of the vortex from the center of the trap. This is the region of thermodynamic stability. 
\section{Vortex dynamics}
The influence of the thermal atoms on the motion of the vortex is the subject of the remainder of this paper, and we will use well established techniques to analyse the observable finite temperature effects on the vortex dynamics and its lifetime.
\subsection{Equations of motion}
As shown in Ref.~\cite{Duine2002}, using techniques developed in Ref.~\cite{Duine2004}, the equation of motion for the vortex coordinates can be mapped to a pair of equivalent Langevin field equations. Writing $z(t)=u_x(t)+iu_y(t)$ for the vortex position in the complex plane, this takes the form of a single equation of motion
\EQ{\label{eq:vlang}
\frac{dz}{dt}=(i\omega_p+\gamma)z+\eta(t)
}
where the complex Gaussian noise is determined by the non-zero correlator
\EQ{\label{eq:vnoise1}
\langle \eta^*(s)\eta(t)\rangle=2\sigma\delta(s-t),
}
and the magnitude of the thermal noise is determined by
\EQ{\label{eq:sig}
\sigma = \frac{q^2i\Sigma^{\rm K}}{4N_c}.
}
For a vortex initially situated at $z(0)=z_0$, the solution
\EQ{
z(t)=z_0e^{(i\omega_p+\gamma)t}+\int_0^t ds\; \eta(s)e^{(i\omega_p+\gamma)(t-s)},
}
together with Eq.~\ref{eq:vnoise1}
can be used to calculate all observables associated with the vortex motion.
\par
 We can also find an equivalent Fokker-Planck equation (FPE) for the Langevin field equation derived above. Solving for the vortex motion is then equivalent to finding the probability distribution for the vortex position in the lab frame, which, from Eq.~\ref{eq:vlang}, can be seen to evolve according to the FPE
\EQ{\label{eq:vFPE}
\frac{\partial P(z,z^*,t)}{\partial t}=\left\{\frac{\partial}{\partial z}(-i\omega_p-\gamma)z+\frac{\partial}{\partial z^*}(i\omega_p-\gamma)z^*+2\sigma\frac{\partial^2}{\partial z\partial z^*}\right\}P(z,z^*,t).
}
This equation is closely related to the FPE for the P-function describing a quantised light field in a damped optical resonator~\cite{Carmichael1999}. Indeed, the FPE is equivalent to the evolution of a damped Boson mode with frequency $\omega_p$, described by density matrix $\rho$, according to the master equation 
\EQ{
\dot{\rho}&=&-i\omega_p[a^\dag a,\rho]+(\sigma-\gamma)(2a\rho a^\dag-a^\dag a\rho-\rho a^\dag a)\nonumber\\
&&+\sigma(2a^\dag\rho a-a a^\dag\rho-\rho  aa^\dag),
}
where $[a,a^\dag]=1$, and $a^\dag$ creates a particle in the oscillator mode with frequency $\omega_p$. The system is analogous to a {\em one} dimensional harmonic oscillator with a peculiar reservoir coupling, provided we regard $u_x$ and $u_y$ as conjugate phase space variables, analogous to the ordinary position and momentum. We will show below that the form of the coupling leads naturally to the introduction of an associated quasimode which neatly encapsulates the concept of vortex stability.
The solution of Eq.~\ref{eq:vFPE} for a vortex initially located at $z_0$ is given by
\EQ{\label{eq:Pdist}
P(z,z^*,t|z_0,z_0^*,0)=\frac{\gamma}{\pi\sigma(e^{2\gamma t}-1)}\exp{\left(-\frac{\gamma|z-z_0 e^{(i\omega_p+\gamma)t}|^2}{\sigma(e^{2\gamma t}-1)}\right)}
}
which provides a complete description of the vortex motion for both thermodynamically stable and unstable regimes.
The mean and variance of the distribution determine all higher order moments, and are given by
\EQ{\label{eq:meanr}
\langle z(t)\rangle&=&z_0e^{(i\omega_p+\gamma)t},\\\label{eq:varr}
\langle z^*(t)z(t)\rangle-\langle z^*(t)\rangle\langle z(t)\rangle&=&\frac{\sigma}{\gamma}(e^{2\gamma t}-1).
}
When the thermal cloud rotates slower than the vortex precesses the dissipation causes the mean radius to increase with time, indicating an instability, whereas if it rotates faster than the critical frequency it is damped with time, indicating stability. In both regimes thermal fluctuations cause the distribution to spread, so that the actual physical stability must be determined by including both processes. The sign of $\gamma$ determines the {\em thermodynamic} stability of the vortex, since this is a consequence of the sign of the precession frequency in the frame of the thermal cloud. As we will see below, thermal fluctuations weaken the thermodynamic stability, so that the vortex always has a finite lifetime. This is a consequence of diffusive instability which means that the the vortex can always reach the edge of the condensate in a finite time.
\subsection{Vortex Quasimode}
It is useful here to consider the angular momentum and energy associated with the vortex. 
We first consider the angular momentum corresponding to the Gaussian ansatz of Eq.~\ref{eq:Gaussian}, which is
\EQ{
\langle L_z\rangle=-i\hbar\int d\mbf{x}\; \phi^*(\mbf{x})(x\partial_y-y\partial_x)\phi(\mbf{x})=\hbar N_c\left(1-\frac{u^2_x(t)+u^2_y(t)}{q^2}\right),
}
so that
\EQ{
N_{\rm v}(t)=N_c\left(1-\frac{u^2_x(t)+u^2_y(t)}{q^2}\right)
}
is understood to be the number of particles contributing to the state with a vortex at radius $[u^2_x(t)+u^2_y(t)]^{1/2}$. We will also find it useful to introduce the number of particles associated with the motion of the vortex core, defined as
\EQ{
n_{\rm v}(t)=N_c-N_{\rm v}(t).
}
When the vortex is stationary at the center of the trap {\em all} atoms are in the vortex state, and the occupation of the mode associated with the vortex position is zero.
\par
To obtain an expression for the energy associated with the position of the vortex core we define the vortex Hamiltonian 
\EQ{\label{eq:vhamil}
H_{\rm v}=n_{\rm v}\hbar\omega_{\rm v}
}
where 
\EQ{
\omega_{\rm v}&=&\Omega-\omega_p
}
is the effective frequency of the mode.
The average vortex energy is $E_{\rm v}\equiv\langle H_{\rm v}\rangle$, or
\EQ{
E_{\rm v}(t)&=&\langle n_{\rm v}(t)\rangle\hbar \omega_{\rm v}.
}
We can regard $\langle n_{\rm v}(t)\rangle$ as the mean occupation of the quasimode, and we will show that this interpretation is consistent with the equipartion theorem. 
For convenience we have chosen the origin of the vortex energy scale as $\hbar\Omega=\hbar\omega_p$, corresponding to the rotation rate of the cloud above which the vortex becomes thermodynamically stable. Note that the absence of a zero point energy term in Eq.~\ref{eq:vhamil} is a consequence of working in the high temperature regime. 
\subsection{Vortex Lifetime I: Diffusion Time}\label{sec:diffinst}
The vortex diffusion process is most easily characterised by the width the position distribution. It is convenient to express the analysis of this section in terms of the {\em rms radius} $\bar{r}\equiv\sqrt{\langle |z|^2\rangle}$, and we will use the notation $N_{\rm v}(\bar{r})=N_c(1-\bar{r}^2/q^2)$ to denote the mean occupation of the vortex state corresponding to the rms-radius $\bar{r}$, and similarly for $E_{\rm v}(\bar{r})$ and $n_{\rm v}(\bar{r})$. We emphasise here that by using $\sqrt{\langle |z|^2\rangle}$ rather than $\langle|z|\rangle$ in our description of the lifetime we will avoid the need to introduce the approximation $\langle 1/|z|\rangle\approx1/\langle |z|\rangle$, which was implicitly used in the calculation of the vortex lifetime in Ref.~\cite{Duine2004}, effectively underestimating the role of fluctuations~\cite{Rapprox}. 
\par
We can now find an estimate for the vortex lifetime.
The equation of motion for the rms radius
\EQ{
\bar{r}(t)&=&\sqrt{|z_0|^2e^{2\gamma t}+\sigma(e^{2\gamma t}-1)/\gamma},
}
takes the form
\EQ{\label{eq:Rrms}
\frac{d\bar{r}(t)}{dt}=\gamma \bar{r}(t)+\frac{\sigma}{\bar{r}(t)}.
}
We can now find the time taken for $\bar{r}(t)$ to change from $\bar{r}_1(t_1)$ to $\bar{r}_2(t_2)$ as
\EQ{\label{eq:lifetime}
\tau(\bar{r}_1,\bar{r}_2) = \int_{\bar{r}_1}^{\bar{r}_2}d\bar{r} \left(\frac{d\bar{r}(t)}{dt}\right)^{-1}=\frac{1}{2\gamma}\ln{\left[\frac{\bar{r}_2^2+\sigma/\gamma}{\bar{r}_1^2+\sigma/\gamma}\right]}.
}
Using Eqs. \ref{eq:gam}, \ref{eq:sig}, and \ref{eq:lifetime} we obtain
\EQ{\label{eq:vlife}
\tau(\bar{r}_1,\bar{r}_2) =\frac{\tau_c}{\beta E_{\rm v}(q)}\ln{\left[\frac{1-\beta E_{\rm v}(\bar{r}_1)}{1-\beta E_{\rm v}(\bar{r}_2)}\right]}.
}
In the above expression we have introduced $\tau_c$ representing the diffusion time obtained by choosing $\bar{r}_1=0$, $\bar{r}_2=q$, and $\Omega=\omega_p$, which takes the form
\EQ{\label{eq:tc}
\tau_c \equiv \tau(0,q)\Big|_{\Omega=\omega_p}=\left(\frac{\hbar}{k_BT}\right)^2\frac{\pi\hbar N_c}{24ma^2}.
}
This universal {\em critical diffusion time} arises when the cloud rotates at the critical frequency for thermodynamic stability of the vortex, and its form is independent of the size of the condensate and the vortex precession frequency, arising only from thermal fluctuations which cause the vortex position to diffuse in time. This is the irreducible timescale associated with the diffusion instability.
We note that the form of Eq.~\ref{eq:vlife} dictates that
$
\bar{r}_1<\bar{r}_2
$
in order to give a well defined positive value. 
\par
\begin{figure}[!htb]
\includegraphics[width=1\columnwidth]{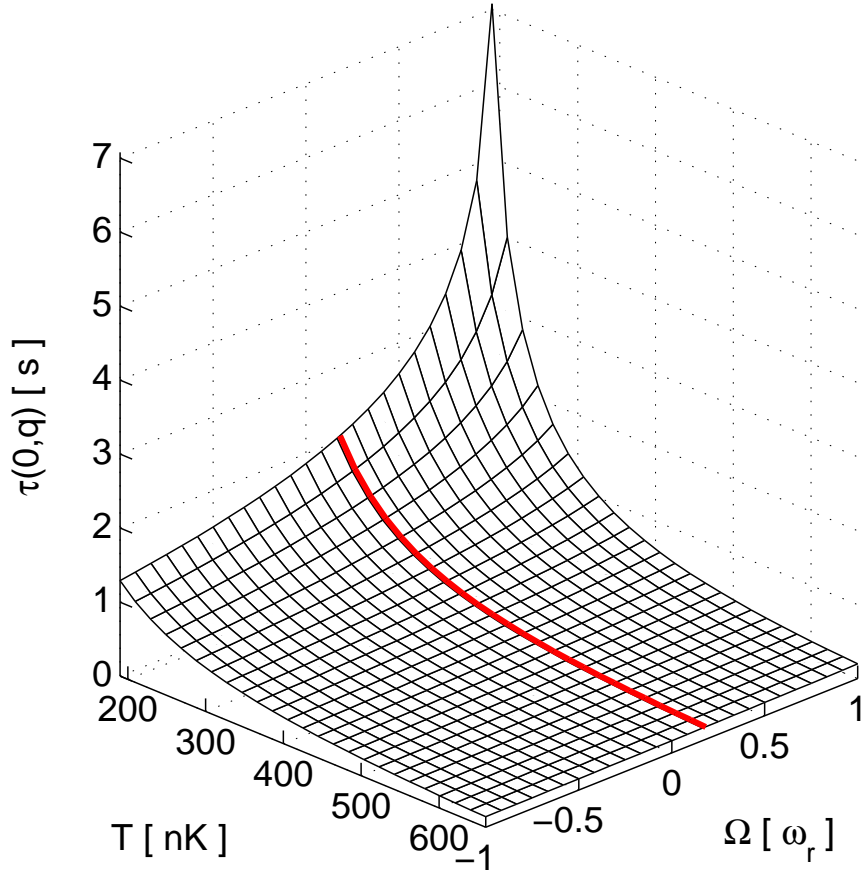}
\caption{Vortex diffusion time for $N_c=500$ atoms of $^{87}{\rm Rb}$ in an oblate trap with frequencies $(\omega_z,\omega_r)=2\pi\times(10,100){\rm Hz}$. The diffusion time is shown for the stable range of $\Omega$ ($-\omega_r<\Omega<\omega_r$), and the bold line (color online) shows the critical lifetime $\tau_c$.}
\vspace{-.3cm}
\label{fig:vlifeRb}
\end{figure}

 
The special case of experimental interest is the vortex lifetime associated with the time for a vortex initially located at the center of the trap to reach $\bar{r}=q$.
In Fig.~\ref{fig:vlifeRb} we present this value for $N_c=500$ atoms of $^{87}{\rm Rb}$.  As we noted above, the Gaussian ansatz is appropriate for weak interactions, which, for the species we consider, requires small condensate occupation. The expression for $\tau_c$ is shown by the dashed line, which also marks the boundary of the thermodynamically stable region, while the surface plot depicts the variation of Eq.~\ref{eq:vlife} with the temperature and angular frequency of the thermal cloud. 
\par 
To summarise, the diffusion time that we have computed in this section is expressed in terms of the variance of the vortex distribution (similar to the result of Ref.~\cite{Duine2004}). This approach reveals a characteristic timescale associated with the diffusive instability, and gives a good first estimate of the vortex lifetime. In the next section we give a more rigorous description of the vortex lifetime using the theory of {\em mean first exit times}~\cite{SM}.
\begin{figure}[!htb]
\includegraphics[width=1\columnwidth]{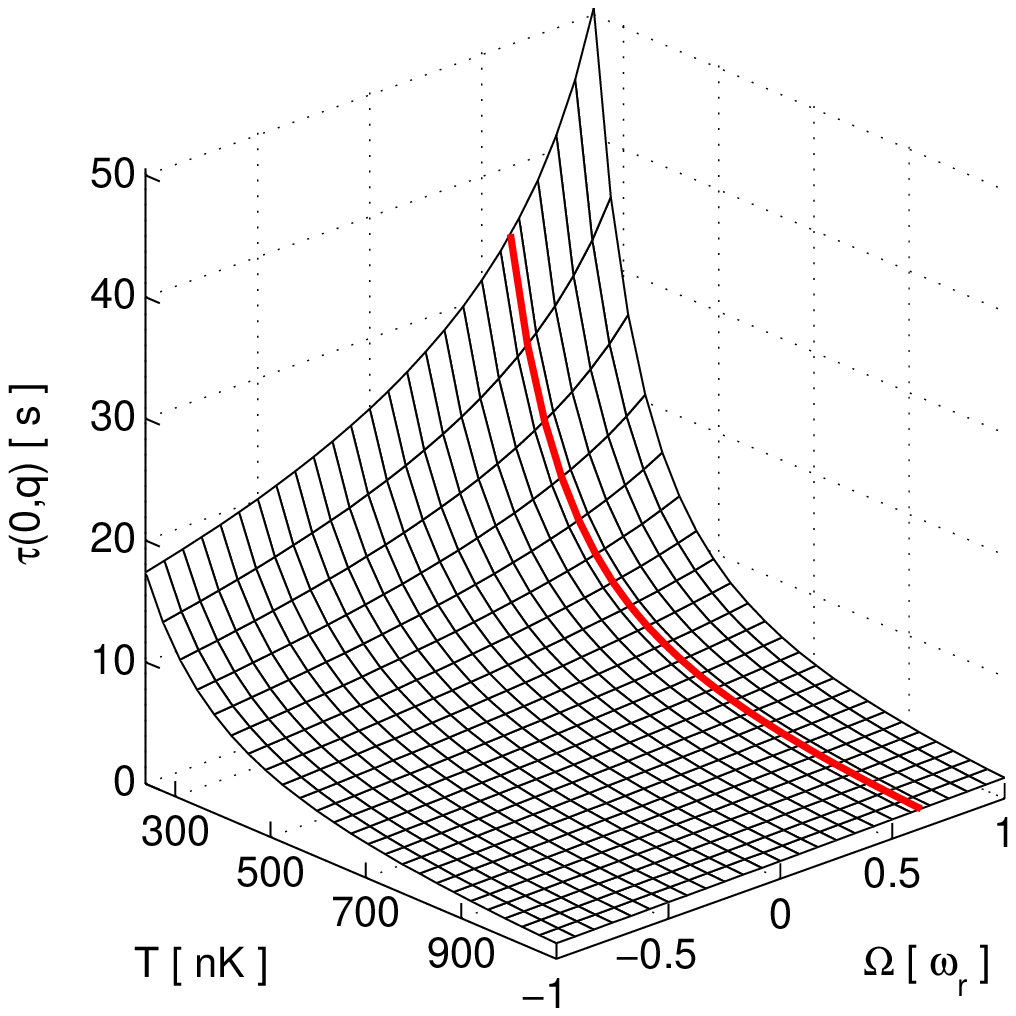}
\caption{Vortex diffusion time for $N_c=800$ atoms of $^{23}{\rm Na}$. The other parameters are the same as those used in Fig. \ref{fig:vlifeRb}}
\vspace{-.3cm}
\label{fig:vlifeNa}
\end{figure}
\subsection{Vortex Lifetime II: Mean Exit Time}
We want to know the mean exit time associated with the escape of the vortex from the region $r=(u_x^2+u_y^2)^{1/2}\leq r_2$, given its initial position $r=r_1$.
 After going into a rotating frame at frequency $\omega_p$, we may rewrite Eq.~\ref{eq:vlang} as the Langevin equation
\EQ{
\frac{dr}{dt} =\gamma r+\frac{\sigma}{2r}+\zeta(t),
}
where the real Gaussian noise is defined by the correlation
\EQ{
\langle \zeta(s)\zeta(t)\rangle=\sigma\delta(s-t).
}
The symmetry of the system means that the vortex diffusion is an effectively one dimensional process in the radial coordinate. The problem of the mean exit time from the region $(0,r_2)$ is formulated in terms of the probability that the vortex is in the region $(0,r_2)$ at time $t$
\EQ{
G(r_1,r_2,t)\equiv\int_0^{r_2} dr^\prime\; p(r^\prime,t\;|r_1,0).
}
Let the time that the particle leaves $(0,r_2)$ be $T.$ Then ${\rm Prob}(T\geq t)=G(r_1,r_2,t)$, and the {\em mean first passage time} is
\EQ{
T(r_1,r_2)=-\int_0^\infty t\; dG(r_1,r_2,t).
}
The solution of this problem is known, and we refer the reader to Ref.~\cite{SM} for the details~\cite{exitSM}. The mean first exit time for the vortex initially at radius $r_1$ to exit from the region $r \leq r_2$ is
\EQ{
T(r_1,r_2)=\frac{\tau_c}{2\beta H_{\rm v}(q)}\ln{\left[\frac{H_{\rm v}(r_1)e^{E_1(-\beta H_{\rm v}(r_1))}}{H_{\rm v}(r_2)e^{E_1(-\beta H_{\rm v}(r_2))}}\right]},
}
where we have made use of the {\em exponential integral}~\cite{Abram}, defined as
\EQ{
E_1(x)=\int_{x}^\infty dt\;\frac{e^{-t}}{t}.
}
Note that the singularity plays no role since the difference of these integrals is always over an interval which excludes the origin~\cite{expint}.
The mean exit time has similar order of magnitude to the expression found in section~\ref{sec:diffinst}, given by Eq.~\ref{eq:vlife}. In particular, when the cloud and vortex co-rotate we have the {\em critical exit time} from the condensate center to the edge 
\EQ{
T_c\equiv T(0,q)\Big|_{\Omega=\omega_p}=\frac{\tau_c}{2},
}
which is a further manifestation of the diffusive instability.
\par
\begin{figure}[!htb]
\includegraphics[width=1\columnwidth]{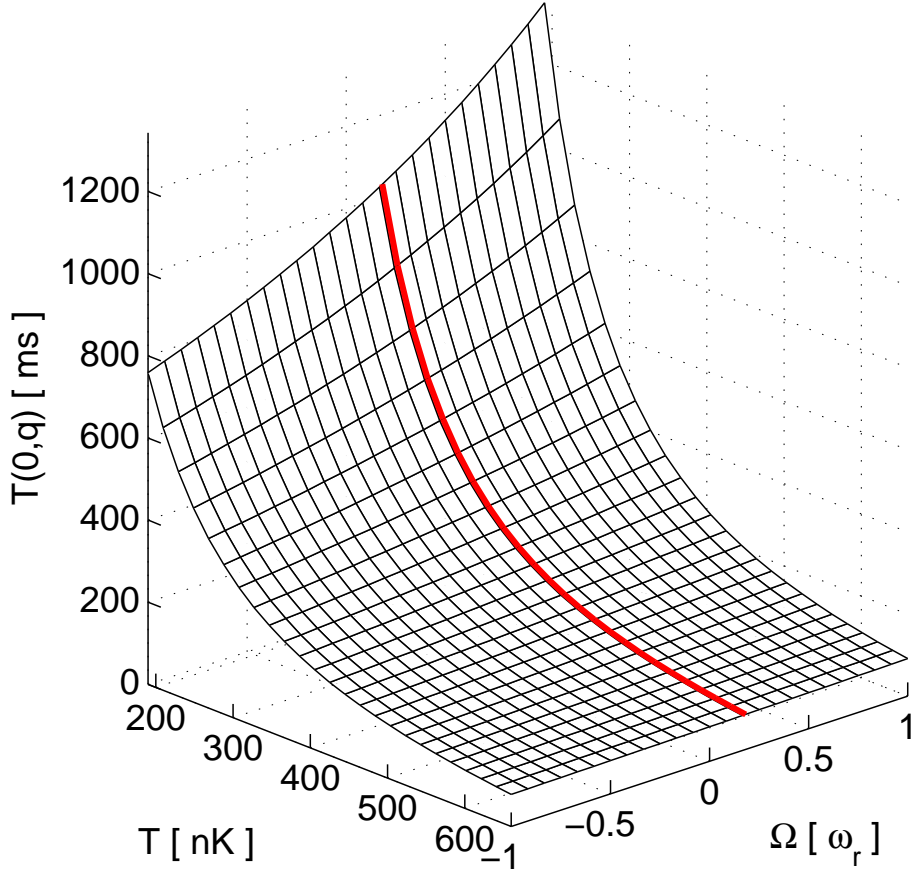}
\caption{Vortex mean exit time for $N_c=500$ atoms of $^{87}{\rm Rb}$ for the same parameters as Fig. \ref{fig:vlifeRb}}
\vspace{-.3cm}
\label{fig:exitRb}
\end{figure}

\begin{figure}[!htb]
\includegraphics[width=1\columnwidth]{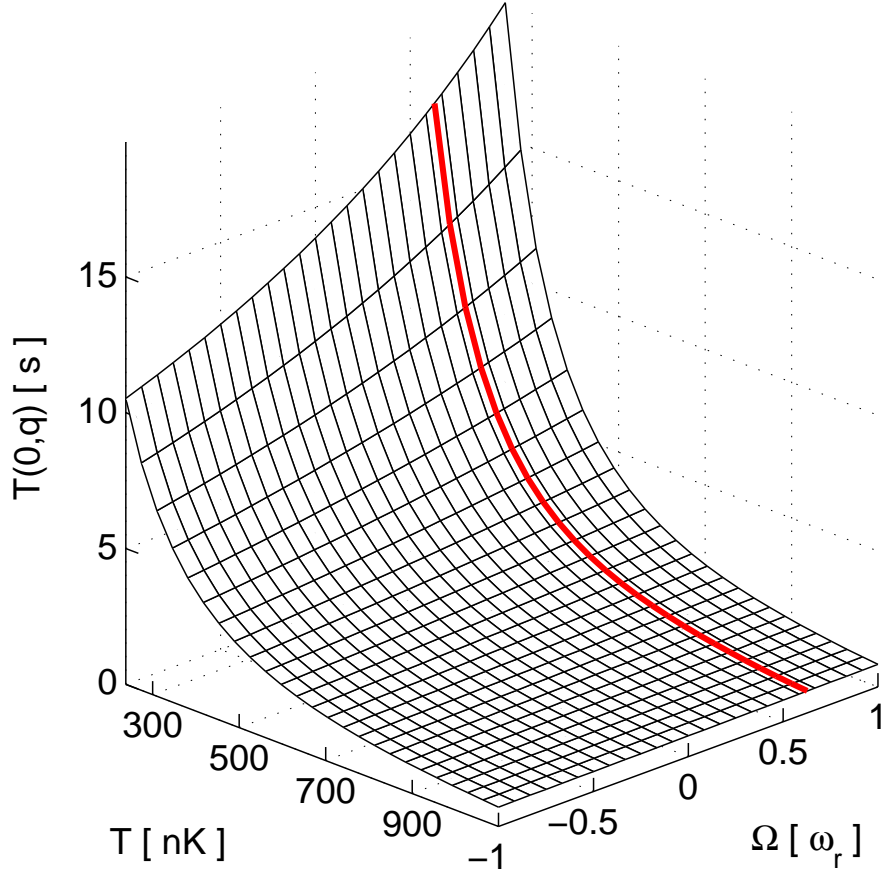}
\caption{Vortex mean exit time for $N_c=800$ atoms of $^{23}{\rm Na}$ for the same parameters as Fig. \ref{fig:vlifeRb}}
\vspace{-.3cm}
\label{fig:exitNa}
\end{figure}

 
In Figs.~\ref{fig:exitRb},~\ref{fig:exitNa} we show the mean exit times for sodium and rubidium for a range of temperatures and angular frequencies of the thermal cloud. As before, we have marked the critical exit time $T_c$ with a dashed line. We note here that as well as being shorter by roughly a factor of three than the diffusion time estimate, the mean exit time has a weaker dependence on the relative motion of cloud and vortex than the diffusion time.
\subsection{Equipartition, Correlations and Fluctuations}
Using Eq.~\ref{eq:vlife} one can find an expression for the time $d\tau$ required for the distribution to evolve from $\bar{r}$ to $\bar{r}+d\bar{r}$. After a little manipulation we obtain the equivalent differential form
\EQ{\label{eq:vplanck}
\frac{d E_{\rm v}(\tau)}{d\tau}=\frac{N_c\hbar\omega_{\rm v}}{\tau_c}\left(1-\beta E_{\rm v}(\tau)\right),
}
which is just the rate equation for the vortex energy. The possibility of  parametrising the energy by the radial diffusion time coordinate $\tau$ arises from the physical fact that the mean energy associated with the vortex only depends on $\bar{r}$. The solution is
\EQ{
E_{\rm v}(\tau_2)=E_{\rm v}(\tau_1)e^{-\beta N_c\hbar\omega_{\rm v}(\tau_2-\tau_1)/\tau_c}+k_B T\left(1-e^{-\beta N_c\hbar\omega_{\rm v}(\tau_2-\tau_1)/\tau_c}\right).
}
When $\omega_{\rm v}>0$, corresponding to the thermodynamically stable scenario, we obtain the steady state occupation of the vortex quasimode
\EQ{\label{eq:nv}
n_{\rm v}(\bar{r})=\frac{k_BT}{\hbar\omega_{\rm v}},
}
in accordance with the equipartition theorem. As expected, the steady state distribution $P_s(z,z^*)$, found from Eq.~\ref{eq:Pdist}, has the form
\EQ{
P_{s}(z,z^*)=\frac{N_c\hbar\omega_{\rm v}}{\pi k_BTq^2}\exp{\left(-\frac{N_c\hbar\omega_{\rm v}}{k_BT}\frac{|z|^2}{q^2}\right)},
}
which is consistent with the equipartition result of Eq.~\ref{eq:nv}. 
\par
Since we have such a simple description, we should also consider the measurable correlation functions for the vortex position that can give information about the thermal statistics of the condensate. It is convenient to discuss these quantities in the regime of thermodynamic stability, where two-time correlation functions reach a steady state behaviour. For the first and second order stationary two time correlation functions we have
\EQ{
g^{(1)}(\tau)&=&{\rm lim}_{t\to \infty}\frac{\langle z^*(t)z(t+\tau)\rangle}{\sqrt{\langle z^*(t)z(t)\rangle\langle z^*(t+\tau)z(t+\tau)\rangle}},\\
g^{(2)}(\tau)&=&{\rm lim}_{t\to \infty}\frac{\langle z^*(t)z^*(t+\tau)z(t+\tau)z(t)\rangle}{\langle z^*(t)z(t)\rangle\langle z^*(t+\tau)z(t+\tau)\rangle}.
}
The correlation functions associated with the steady state vortex motion are
\EQ{\label{eq:g1}
g^{(1)}(\tau)&=&e^{(-|\gamma|+i\omega_p)\tau},\\
\label{eq:g2}
g^{(2)}(\tau)&=&1+e^{-2|\gamma|\tau}.
}
The Fourier transform of Eq.~\ref{eq:g1} gives a Lorentzian power spectrum associated with the vortex postion. From Eq.~\ref{eq:g2} we see that there is a `thermal bunching' effect~\cite{Carmichael1999} for the variance, corresponding to the increased probability of vortex-vortex coincidence at short times. Eq.~\ref{eq:g2} also tells us about the steady state spectrum of energy fluctuations of the vortex quasimode, since we may rewrite this as
\EQ{\label{eq:Hvspec}
\frac{\langle H_{\rm v}(0)H_{\rm v}(\tau)\rangle-\langle H_{\rm v}(0)\rangle \langle H_{\rm v}(\tau)\rangle}{\langle H_{\rm v}(0)\rangle \langle H_{\rm v}(\tau)\rangle}=e^{-2|\gamma|\tau},
}
giving a broad Lorentzian spectrum for the energy fluctuations arising from collisions with the thermal cloud. The power spectrum potentially has an important role in setting the thermal fluctuations arising in the condensate due to the vortex, since the vortex effectively acts as a radiating oscillator with this energy spectrum. However, the process of energy transmission into the condensate from a jostling vortex is expected to depend on the details of the vortex core structure, which is beyond the scope of this paper.
\par
Finally, we remark that the lifetime of the vortex can be estimated by short time measurements of the mean and variance of the vortex location. The radial drift rate is determined by $\gamma$ in Eq.~\ref{eq:meanr}, while from Eq.~\ref{eq:varr} we see that the radial variance evolves as ${\rm Var}(r)\approx 2\sigma t$ for short times, exactly as for the motion of a Brownian particle suspended in a viscous fluid. Since these two quantities, along with the vortex precession frequency and the angular frequency of the cloud, completely determine the vortex evolution, it is a simple matter to estimate the lifetime using Eq.~\ref{eq:vlife} once these are known.
\section{Conclusion}
We have shown that a perturbative description which includes the rotation of the thermal cloud can be developed using the stochastic variational method of Duine {\em et al}~\cite{Duine2002,Duine2004}. The vortex acts as a probe for thermal fluctuations, with the lifetime depending  on both the thermodynamic and diffusive instabilities through the temperature and motion of the thermal cloud. The process of vortex exit from the condensate can be satisfactorily described using the theory of mean exit times, and the theory predicts a shorter lifetime for the vortex than the diffusion time method. Nevertheless, the diffusive effect of the cloud is much stronger than the damping effect; in particular, if one neglects the noise from the description the exit time predictions are of the order of tens of diffusion times.
\par
The success of the approach relies on the physical fact that the variation of the atom number with vortex radius can be neglected to a first approximation. Even if the corrections arising from changing atom number were included, the method is not expected to give good predictions of the detailed motion near the condensate edge since it only affords a quadratic approximation to the chemical potential. Fortunately the motion in this region does not contribute significantly to the lifetime. Finally, we remark that the lifetime we have calculated bears little resemblance to the result of Fedichev and Shlyapnikov~\cite{Fedichev1999}, apart from a logarithmic behaviour which is characteristic of two dimensional systems. This is not unexpected since their analysis of vortex damping caused by quasiparticles in the Thomas-Fermi regime is for an entirely different scenario to the weakly interacting, thermally damped regime considered here. This raises the possibility of extending the high temperature classical field method beyond weak interactions to the Thomas-Fermi regime, which is a matter for future research.
\par
AB thanks Murray Olsen, Matthew Davis and Peter Drummond for helpful comments.
This work was supported by \TAD, \MF, and the \ARC.

\begin{thebibliography}{10}
\bibitem{expint}
For negative arguments $(\gamma < 0)$ the definition is $E_1(-x\pm
  i0)=-Ei(x)\mp i\pi$, where $Ei(x)=-{\cal P}\int_{-x}^\infty dt \;e^{-t}/t$,
  and ${\cal P}$ denotes the principal value. See Ref.~\cite{Abram}, p 228 for
  details.
\end{thebibliography}

\end{document}